\documentclass[%
 reprint,
superscriptaddress,
 amsmath,amssymb,
 aps,
 prx,
floatfix,
]{revtex4-2}

\usepackage{graphicx}% Include figure files
\usepackage{dcolumn}% Align table columns on decimal point
\usepackage{bm}% bold math
\usepackage{physics}
\usepackage{tikz}
\usepackage{quantikz}
\usetikzlibrary{arrows.meta,positioning,shapes.geometric,fit,calc,backgrounds}
\usepackage{amsmath}
\usepackage{xcolor}
\usepackage{adjustbox}
\usepackage{mathtools}
\usepackage{booktabs}

\definecolor{qft}{RGB}{56,134,210}
\definecolor{phase}{RGB}{220,140,40}
\definecolor{dil}{RGB}{180,60,60}
\definecolor{nh}{RGB}{60,160,90}

\usepackage[colorlinks=true,allcolors=blue]{hyperref}
\usepackage{orcidlink}

\newcommand{\IF}{\mathrm{IF}}
\newcommand{\psiexact}{\ket{\psi_{\mathrm{exact}}}}
\newcommand{\psidmrg}{\ket{\psi_{\mathrm{DMRG}}}}
\newcommand{\psienc}{\ket{\psi_{\mathrm{enc}}}}

\newcommand{\chicut}{\chi_{\mathrm{cut}}}
\newcommand{\chimax}{\chi_{\mathrm{max}}}
\newcommand{\Lstar}{L^{*}}

\newcommand{\dtau}{\Delta \tau}
\newcommand{\eps}{\varepsilon}

\newcommand{\PiteRZtwo}[1]{\ensuremath{R_z(#1)}}
\newcommand{\PiteURTEtwo}[1]{\ensuremath{\hat U_{\mathrm{RTE}}(#1)}}
\newcommand{\PiteRZZtwo}[1]{\ensuremath{R_{ZZ}(#1)}}
\newcommand{\PitecOtwo}[1]{\ensuremath{c\text{-}\hat U_{\mathrm O}(#1)}}
\newcommand{\PitecEtwo}[1]{\ensuremath{c\text{-}\hat U_{\mathrm E}(#1)}}

\begin{document}

\title{Overcoming the Matrix-Product-State Encoding Barrier via DMRG-Guided Probabilistic
Imaginary-Time Evolution}

\author{Masari Watanabe\orcidlink{0000-0003-3186-535X}}
\email{mwatanabe@quemix.com}
\affiliation{Quemix Inc., Taiyo Life Nihonbashi Building, 2-11-2,
Nihonbashi Chuo-ku, Tokyo 103-0027, Japan}
\affiliation{Department of Physics, The University of Tokyo, Hongo, Bunkyo-ku, Tokyo 113-0033, Japan}

\author{Hirofumi Nishi\orcidlink{0000-0001-5155-6605}}
\affiliation{Quemix Inc., Taiyo Life Nihonbashi Building, 2-11-2,
Nihonbashi Chuo-ku, Tokyo 103-0027, Japan}
\affiliation{Department of Physics, The University of Tokyo, Hongo, Bunkyo-ku, Tokyo 113-0033, Japan}

\author{Taichi Kosugi\orcidlink{0000-0003-3379-3361}}
\affiliation{Quemix Inc., Taiyo Life Nihonbashi Building, 2-11-2,
Nihonbashi Chuo-ku, Tokyo 103-0027, Japan}
\affiliation{Department of Physics, The University of Tokyo, Hongo, Bunkyo-ku, Tokyo 113-0033, Japan}

\author{Shinji Tsuneyuki\orcidlink{0009-0004-8790-7429}}
\affiliation{Department of Physics, The University of Tokyo, Hongo, Bunkyo-ku, Tokyo 113-0033, Japan}

\author{Yu-ichiro Matsushita\orcidlink{0000-0002-9254-5918}}
\affiliation{Quemix Inc., Taiyo Life Nihonbashi Building, 2-11-2, Nihonbashi Chuo-ku, Tokyo 103-0027, Japan}
\affiliation{Department of Physics, The University of Tokyo, Hongo, Bunkyo-ku, Tokyo 113-0033, Japan}
\affiliation{Quantum Materials and Applications Research Center,
National Institutes for Quantum Science and Technology, Tokyo 152-8550, Japan}

\date{\today}% It is always \today, today,

\begin{abstract}
Ground-state preparation is a fundamental task in quantum simulation, because
the overlap of the prepared state with the true ground state significantly affects
the overall cost of subsequent quantum algorithms.
We propose a three-stage framework in which a matrix product state (MPS) of an
$N$-site system obtained by the density-matrix renormalization group (DMRG) is
loaded onto an $N$-qubit quantum register through an optimization-free matrix
product disentangler (MPD) encoding circuit, and the residual error is then
reduced by probabilistic imaginary-time evolution (PITE).
We demonstrate that the central-bond Schmidt rank of intermediate states during
MPS encoding grows logistically with the number of layers.
Its inflection point $L^{*}$ marks the boundary of the efficient encoding regime.
Beyond this point, the gain in fidelity slows rapidly, and the number of
additional MPD layers required to reach a target infidelity $\varepsilon$ empirically
scales as $\order{N^5\log(N/\varepsilon)}$.
To avoid this encoding-only tail, we stop the encoder at $L^{*}$ and suppress
the remaining excited-state components by PITE, with the linear PITE schedule
fixed deterministically from the ground-state energy, the effective gap, and the
reference overlap estimated by DMRG.
Numerical experiments on the spin-$1/2$ staggered-field Heisenberg chain show
that the framework avoids very deep encoding circuits and substantially suppresses
the post-selection overhead intrinsic to PITE.
In the tested size range, the MPS-derived initial state keeps the cumulative
success probability of order unity, substantially higher than that of the N\'eel
product state, and the expected post-selection-weighted circuit cost to reach
the target accuracy is reduced accordingly.
Combining classical preprocessing by DMRG, optimization-free MPS encoding, and
deterministically scheduled PITE, the present framework offers a practical hybrid
route to ground-state preparation in quantum simulation.
\end{abstract}

\maketitle

% ============================================================
\section{\label{sec:intro}Introduction}
% ============================================================

Quantum algorithms have been actively developed across quantum chemistry, quantum many-body simulation, partial differential equations, and fluid dynamics~\cite{lloyd1996,cao2019,bauer2020,yoshioka2024,childs2021pde,bharadwaj2023}.
In these applications, ground-state energies, low-energy excitations, and correlation functions determine many of the physical and chemical properties of interest, and ground-state computation therefore plays a central role~\cite{aspuruguzik2005}.
The overlap of the prepared state with the true ground state significantly affects the overall cost of the subsequent algorithm, including the success probability, the number of measurements, and the post-selection overhead~\cite{lee2023,fomichev2024,ollitrault2024,berry2025}.
Ground-state preparation is therefore a central component of practical quantum algorithms.

Encoding circuits inspired by matrix product states (MPSs) form one promising class of state-preparation strategies.
Sch\"on \textit{et al.}\ showed that an MPS of finite bond dimension can be generated exactly through sequential interactions with an auxiliary system~\cite{schon2005,schon2007}.
On actual quantum devices, however, the direct implementation of multi-qubit unitaries corresponding to large auxiliary spaces or high bond dimensions is prohibitively expensive, and practical encoders compile a given MPS into one- and two-qubit gates with a controlled approximation.
Ran's sequential encoding constructs each layer from a low-bond-dimension MPS and applies the corresponding two-qubit gates iteratively, thereby converting an MPS into a quantum circuit without variational optimization~\cite{ran2020}.
More recently, Malz \textit{et al.}\ proved that the optimal unitary-circuit depth for translation-invariant normal MPS is $\Omega(\log N)$ and gave an algorithm that saturates this bound~\cite{malz2024}, and Smith \textit{et al.}\ showed that adaptive circuits with mid-circuit measurements and classical feedforward can achieve constant depth~\cite{smith2024}.
Beyond the translation invariance and fixed-bond-dimension assumptions underlying these optimal-depth results, the conditions under which simpler, near-term-friendly sequential MPS encoders operate efficiently remain insufficiently understood.

The density-matrix renormalization group (DMRG) is the standard classical method for efficiently obtaining approximate ground states in MPS form, and enables high-accuracy ground-state computations across a wide range of problems from quantum chemistry to many-body simulation~\cite{white1992,white1993,schollwoeck2011}.
In this work we encode a DMRG-derived MPS into a quantum circuit through the Ran-type sequential encoder.
Because a DMRG-MPS has low entanglement by construction, the resulting quantum state can serve as a high-quality initial state for subsequent quantum algorithms.

Ground-state calculation on quantum devices is broadly approached either by variational quantum algorithms or by nonvariational projection- and filtering-type methods.
The former, represented by the variational quantum eigensolver (VQE)~\cite{peruzzo2014}, use shallow circuits but rely on classical optimization with repeated measurements, which can become the rate-limiting step~\cite{cerezo2021,mcclean2018}.
Within the latter class, block-encoding and quantum signal processing enable near-optimal black-box spectral filtering for ground-state preparation, provided that one has access to a Hamiltonian block-encoding and an initial-state preparation oracle with nontrivial ground-state overlap~\cite{lintong2020}.
These results give strong query-complexity guarantees, but leave open the complementary implementation-level question of how such a high-overlap initial state should be constructed from the structure of a target many-body system.
Among more direct real-time-evolution-based nonvariational methods, probabilistic imaginary-time evolution (PITE) implements imaginary-time evolution stochastically using a single ancilla qubit together with forward and backward real-time evolutions~\cite{kosugi2022}.
For this family of probabilistic filtering methods, scheduling the time-evolution parameters has been shown to improve convergence~\cite{nishi2023optimal}.
Although PITE does not require classical variational optimization, several implementation-dependent choices remain, including the schedule, the energy shift, the number of steps, and the Trotter decomposition.

\begin{figure*}[t]
\centering
  \includegraphics[width=0.92\textwidth]{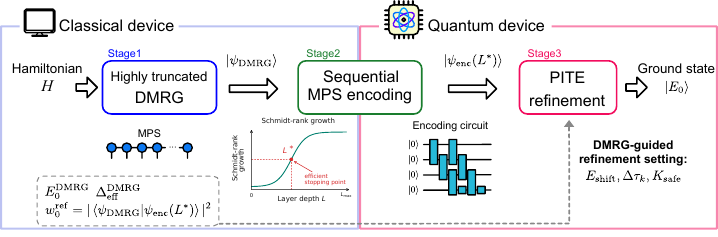}

\caption[Three-stage workflow]{
\textbf{Proposed three-stage workflow.}
DMRG provides an MPS and low-energy information as classical inputs. Sequential MPS encoding then prepares a quantum initial state up to $\Lstar$ layers. The ground-state component is refined by deterministically scheduled PITE, whose schedule uses the approximate ground-state energy, effective gap, and reference overlap from DMRG.
}
\label{fig:framework_and_pite}
\end{figure*}

Here, we propose an optimization-free three-stage workflow for ground-state preparation that integrates DMRG, MPS encoding, and PITE refinement (Fig.~\ref{fig:framework_and_pite}).
We first transfer the DMRG-MPS to a quantum register through the sequential encoder and analyze the encoding process via the growth of the central-bond Schmidt rank.
This growth follows a logistic curve in the number of layers, whose inflection point $\Lstar$ provides a natural boundary of the efficient encoding regime. Beyond $\Lstar$, the additional encoding depth required to reduce the infidelity scales empirically as $\order{N^5\log(N/\eps)}$, and an encoding-only strategy rapidly becomes inefficient.
We therefore stop the encoder at $\Lstar$ and let PITE recover the residual, with the linear PITE schedule fixed deterministically from the DMRG ground-state energy, the effective gap, and the reference overlap.
For the spin-1/2 staggered-field Heisenberg chain, the MPS-encoded initial state keeps the cumulative success probability of order unity, substantially higher than that of the N\'eel product state, and thereby substantially reduces the post-selection-weighted effective cost.

% ============================================================
\section{\label{sec:model}Three-Stage Framework for Ground-State Preparation}
% ============================================================
For a target Hamiltonian $H$, the proposed workflow consists of three stages:
(i) obtaining an MPS initial state that approximates the ground state by DMRG,
(ii) transferring it to a quantum circuit by sequential MPS encoding, and
(iii) refining the ground-state component by PITE.
The structure is illustrated in Fig.~\ref{fig:framework_and_pite}.
We propose to truncate the encoder of Stage 2 at $\Lstar$ layers and feed the resulting state directly to the PITE stage; we show below that this is an efficient route to ground-state preparation.

\subsection{\label{sec:dmrg_mps}Matrix product states and DMRG}

An $N$-site quantum state $\ket{\psi}$ admits an MPS representation with bond dimension $\chi$,
\begin{equation}
  \ket{\psi} = \sum_{\sigma_1,\ldots,\sigma_N}
  A^{[1]\sigma_1} A^{[2]\sigma_2} \cdots A^{[N]\sigma_N}
  \ket{\sigma_1\sigma_2\cdots\sigma_N} ,
  \label{eq:mps}
\end{equation}
where $A^{[n]\sigma_n}$ is a $D_{n-1}\times D_n$ matrix at site $n$, $\sigma_n\in\{0,1\}$ is the physical index, and $D_0=D_N=1$ are imposed as boundary conditions. The maximum bond dimension $\chi=\max_n D_n$ controls the expressive power of the MPS. We obtain an approximate ground state $\psidmrg$ by DMRG with the bond dimension truncated to $\chi=N$, and use it as the classical input to the quantum-circuit encoder.

The MPS can be cast into left-canonical form, in which
\begin{equation}
  \sum_{\sigma_n,a_n}
  A^{[n]\sigma_n}_{a_{n-1},a_n}
  A^{[n]\sigma_n *}_{a'_{n-1},a_n}
  =\delta_{a_{n-1},a'_{n-1}} .
  \label{eq:left_canonical}
\end{equation}
Equation~\eqref{eq:left_canonical} states that each tensor $A^{[n]\sigma_n}$ is an isometry from $a_{n-1}$ to $(\sigma_n,a_n)$. Completing this isometry to a unitary gives a single matrix product disentangler (MPD) layer that disentangles the MPS toward a product state.

\subsection{\label{sec:sequential_encoding}Ran-type sequential MPS encoding}

\begin{figure}[t]
  \centering

  \begin{adjustbox}{max width=\linewidth}
  \begin{quantikz}[row sep={0.76cm,between origins}, column sep=0.1cm]
  \lstick{$\ket{0}_1$}
    & \gate[2,style={fill=blue!10}]{\ensuremath{G^{[1]}}}
    & \qw & \qw & \qw & \qw & \qw
      \slice[
        style={red,dash pattern=on 1pt off 2pt,thick},
        label style={red}
      ]{\ensuremath{\hat{U}_2^\dagger}}
    & \qw & \qw & \qw
    & \gate[2,style={fill=orange!12}]{\ensuremath{G^{[1]}}}
    & \qw & \qw & \qw & \qw & \qw
      \slice[
        style={red,dash pattern=on 1pt off 2pt,thick},
        label style={red}
      ]{\ensuremath{\hat{U}_1^\dagger}}
    & \rstick{} \\
  \lstick{$\ket{0}_2$}
    & \qw
    & \gate[2,style={fill=blue!10}]{\ensuremath{G^{[2]}}}
    & \qw & \qw & \qw & \qw
    & \qw & \qw & \qw
    & \qw
    & \gate[2,style={fill=orange!12}]{\ensuremath{G^{[2]}}}
    & \qw & \qw & \qw & \qw
    & \rstick{} \\
  \lstick{$\ket{0}_3$}
    & \qw & \qw
    & \gate[2,style={fill=blue!10}]{\ensuremath{G^{[3]}}}
    & \qw & \qw & \qw
    & \qw & \qw & \qw
    & \qw & \qw
    & \gate[2,style={fill=orange!12}]{\ensuremath{G^{[3]}}}
    & \qw & \qw & \qw
    & \rstick{} \\
  \lstick{$\ket{0}_4$}
    & \qw & \qw & \qw
    & \gate[2,style={fill=blue!10}]{\ensuremath{G^{[4]}}}
    & \qw & \qw
    & \qw & \qw & \qw
    & \qw & \qw & \qw
    & \gate[2,style={fill=orange!12}]{\ensuremath{G^{[4]}}}
    & \qw & \qw
    & \rstick{} \\
  \lstick{$\ket{0}_5$}
    & \qw & \qw & \qw & \qw
    & \gate[2,style={fill=blue!10}]{\ensuremath{G^{[5]}}}
    & \qw
    & \qw & \qw & \qw
    & \qw & \qw & \qw & \qw
    & \gate[2,style={fill=orange!12}]{\ensuremath{G^{[5]}}}
    & \qw
    & \rstick{} \\
  \lstick{$\ket{0}_6$}
    & \qw & \qw & \qw & \qw & \qw
    & \gate[style={fill=blue!10}]{\ensuremath{G^{[6]}}}
    & \qw & \qw & \qw
    & \qw & \qw & \qw & \qw & \qw
    & \gate[style={fill=orange!12}]{\ensuremath{G^{[6]}}}
    & \rstick{}
  \end{quantikz}
  \end{adjustbox}

  \caption[Circuit elements for state preparation]{%
    \textbf{Sequential MPS encoding circuit for $N=6$ and $L=2$.}
    Each MPD layer $\hat U_l$ consists of one single-qubit gate and $N-1$ two-qubit gates.
    The gate labels are ordered from $G^{[1]}$ at the upper-left to $G^{[N]}$ at the lower-right in each layer.
    The vertical wire order follows the MPS site order, with site $1$ at the top and site $N$ at the bottom.
    These gates are constructed directly from the left-canonical MPS tensors by kernel completion, without optimization~\cite{ran2020}.
    For state preparation, the adjoint $U_{\mathrm{enc}}=\hat U_1^\dagger \hat U_2^\dagger$ is applied to $\ket{0}^{\otimes N}$.
  }
  \label{fig:ran_circuit}
\end{figure}
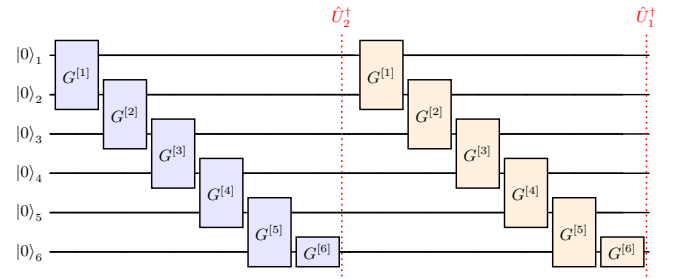

Ran's sequential MPS encoding method~\cite{ran2020} constructs a disentangling circuit that maps a target MPS to the product state $\ket{0}^{\otimes N}$, and uses its inverse as the state-preparation circuit. The corresponding circuit layout is shown in Fig.~\ref{fig:ran_circuit}. At each layer the current MPS $\ket{\psi_k}$ is first optimally truncated to bond dimension $\chi=2$, yielding $\ket{\tilde{\psi}_k}$, and an MPD $\hat U_k$ is then built to map this rank-2 MPS to a product state. The whole procedure is classical. There is no variational optimization performed on the quantum device.

Consider a single layer. The left-canonical MPS tensor $A^{[n]\sigma_n}_{a_{n-1},a_n}$ is an isometry from $a_{n-1}$ to $(\sigma_n,a_n)$. For a rank-2 MPS with physical dimension $d=2$ this isometry can be completed into a $4\times4$ unitary. We pick an orthonormal basis of the kernel orthogonal to the column space of $A^{[n]}$ and define the two-qubit gate $G^{[n]}$ by
\begin{align}
  G^{[n]}_{0,j,k,l} &= A^{[n]}_{j,k,l},
  \label{eq:G_from_A}\\
  \sum_{k,l}G^{[n]*}_{i',j',k,l}G^{[n]}_{i,j,k,l}
  &=\delta_{i'i}\delta_{j'j} ,
  \label{eq:G_unitary}
\end{align}
so that $G^{[n]}$ is unitary. The boundary tensors at $n=1$ and $n=N$ are completed analogously; the $n=N$ gate reduces to a single-qubit unitary.

At iteration $k$ we build the rank-2 approximation $\ket{\tilde{\psi}_k}$ of $\ket{\psi_k}$ and construct the disentangler $\hat U_k$ for $\ket{\tilde{\psi}_k}$. Crucially, $\hat U_k$ is then applied to the original, untruncated state:
\begin{equation}
  \ket{\psi_{k+1}}=\hat U_k\ket{\psi_k} .
  \label{eq:ran_update}
\end{equation}
Because $\hat U_k$ is constructed only for $\ket{\tilde{\psi}_k}$, it is generally not an optimal disentangler for the full $\ket{\psi_k}$. Components beyond the rank-2 truncation undergo a nontrivial unitary rotation. This built-in nonoptimality is the source of the Schmidt-rank growth analyzed in Sec.~\ref{sec:encoding} and of the loss of encoding efficiency seen in deep MPD circuits.

The $L$-layer encoding circuit is
\begin{equation}
  U_{\mathrm{enc}}(L)=\hat U_1^\dagger \hat U_2^\dagger\cdots \hat U_L^\dagger
  \label{eq:U_enc}
\end{equation}
and the prepared quantum state is
\begin{equation}
  \psienc(L)=U_{\mathrm{enc}}(L)\ket{0}^{\otimes N}.
  \label{eq:psi_enc}
\end{equation}

\subsection{\label{sec:pite_overview}PITE overview}

PITE realizes the nonunitary operator $M=e^{-H\tau}$ as a unitary acting on an enlarged Hilbert space with a single ancilla, and produces the imaginary-time-evolved state stochastically by post-selecting on the ancilla outcome.

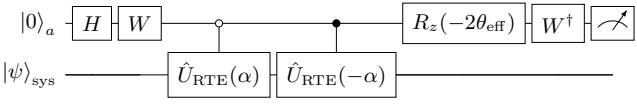
\begin{figure}[hbt]
  \centering
  \resizebox{\linewidth}{!}{\begin{quantikz}[
  row sep={0.76cm,between origins},
  column sep=0.1cm,
  thin lines
]
\lstick{$\ket{0}_a$}
& \gate{H}
& \gate{W}
& \octrl{1}
& \ctrl{1}
& \gate{\PiteRZtwo{-2\theta_{\mathrm{eff}}}}
& \gate{W^\dagger}
& \meter{}
\\
\lstick{$\ket{\psi}_{\mathrm{sys}}$}
& \qw
& \qw
& \gate{\PiteURTEtwo{\alpha}}
& \gate{\PiteURTEtwo{-\alpha}}
& \qw
& \qw
& \qw
\end{quantikz}
}
  \caption[Circuit elements for state preparation]{%
    \textbf{One step of first-order PITE.}
    Post-selecting the ancilla measurement on $\ket{0}_a$, the forward and backward real-time evolutions $\hat U_{\mathrm{RTE}}(\pm\alpha)$ together with the ancilla rotations implement a single stochastic step of imaginary-time evolution on the system register. The $W$ gate, the $R_z(-2\theta_{\rm eff})$ rotation, and the RZZ-native realization of $\hat U_{\rm RTE}$ are detailed in Appendix~\ref{app:rzz_resource}.
  }
  \label{fig:pite_circuit}
\end{figure}

In first-order PITE, the nonunitary operator $M_k=e^{-H\dtau_k}$ is approximated to first order in the imaginary-time step $\dtau_k$ by a filter built from real-time-evolution operators,
\begin{equation}
  M_k
  \simeq
  \gamma_k
  \left[
  \cos(Hs_k\dtau_k)
  -
  \frac{1}{s_k}\sin(Hs_k\dtau_k)
  \right]
  \equiv f_k(H) ,
  \label{eq:pite_filter}
\end{equation}
where $s_k=\gamma_k/\sqrt{1-\gamma_k^2}$. A schematic circuit for one step is shown in Fig.~\ref{fig:pite_circuit}. After $K$ steps,
\begin{equation}
  \ket{\Psi_K}
  =
  \frac{1}{\sqrt{P_K}}
  \prod_{k=1}^{K} f_k(H)\ket{\psi_0} ,
  \label{eq:pite_state}
\end{equation}
with cumulative success probability
\begin{equation}
  P_K=\mel{\psi_0}{\left[\prod_{k=1}^{K} f_k(H)\right]^2}{\psi_0} .
\end{equation}
The total cost of PITE depends sensitively on the initial ground-state overlap $|c_0|^2$. The quality of the MPS-encoded state from Stage 2 therefore determines the effective cost of Stage 3.

% ============================================================
\section{\label{sec:encoding}Schmidt-Rank Growth and Scrambling-Like Encoding Barrier}
% ============================================================
\subsection{\label{sec:target_model}Target model}

For benchmark calculations, we use a spin-1/2 one-dimensional antiferromagnetic Heisenberg spin chain in a staggered magnetic field. The Hamiltonian is described as
\begin{equation}
  H
  = J \sum_{i=1}^{N-1} \bm{S}_i\cdot \bm{S}_{i+1}
    + \sum_{i=1}^{N} h_z (-1)^i S_i^z ,
  \label{eq:ham_spin}
\end{equation}
where $\bm{S}_i=\bm{\sigma}_i/2$, $J>0$ is the antiferromagnetic exchange coupling, and $h_z\ge 0$ is the amplitude of the staggered field. We adopt open boundary conditions (OBCs) and set $J=1$ as the energy unit throughout. In Pauli form,
\begin{equation}
  H
  = \frac{1}{4}\sum_{i=1}^{N-1}
  \left(
  \sigma_i^x\sigma_{i+1}^x+
  \sigma_i^y\sigma_{i+1}^y+
  \sigma_i^z\sigma_{i+1}^z
  \right)
  +\frac{h_z}{2}\sum_{i=1}^{N}(-1)^i\sigma_i^z .
  \label{eq:ham_pauli}
\end{equation}
At $h_z=0$ the system is gapless with entanglement entropy growing logarithmically with subsystem size. For $h_z>0$ the staggered field selects a N\'eel-ordered ground state and opens a finite gap $\Delta=E_1-E_0$. By varying $h_z$, we can therefore systematically examine the relationship between the gap, the correlation length, and the encoding efficiency.

For a quantum state $\ket{\psi}$ and the exact ground state $\psiexact$, we define the fidelity $F=|\braket{\psi_{\mathrm{exact}}}{\psi}|^2$ and the infidelity $\IF=1-F$, and denote by $\eps$ the target accuracy on the infidelity.

\subsection{\label{sec:numerical_implementation}Numerical implementation}

The DMRG ground-state calculations for the Hamiltonian in Eq.~\eqref{eq:ham_spin} were performed using the Python interface of \texttt{block2}~\cite{zhai2023block2}.
Unless otherwise stated, the maximum DMRG bond dimension was set to $\chi=N$.
The resulting MPS tensors were then handled with \texttt{quimb}~\cite{gray2018quimb} for tensor-network post-processing, including canonicalization and Schmidt-spectrum calculations used in the MPS encoding analysis.

\subsection{\label{sec:encoding_phenomenology}Schmidt rank growth under disentangling}

\begin{figure*}[t]
  \centering
  \includegraphics[width=0.92\textwidth]{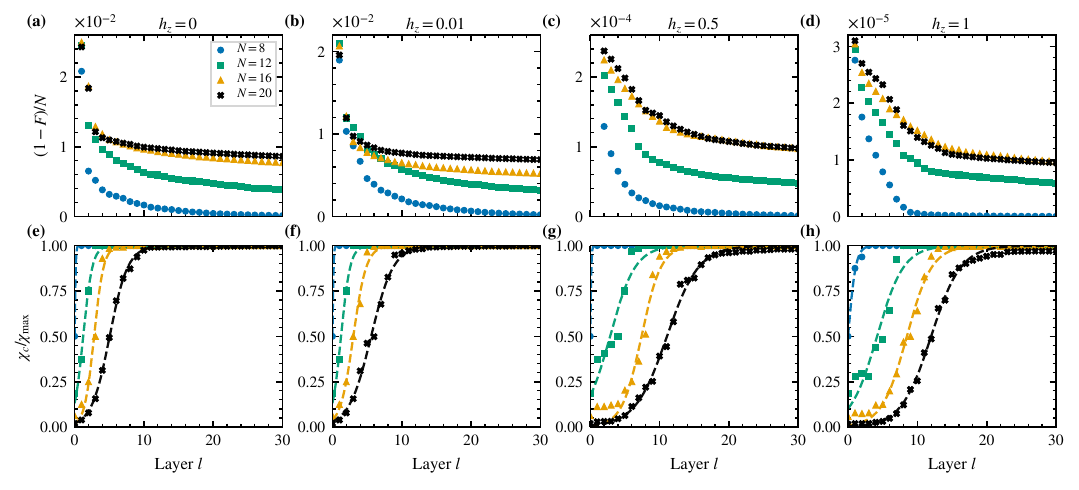}
  \caption[Encoding infidelity and Schmidt-rank growth]{
  \textbf{Layer dependence of the per-site infidelity $\IF(l)/N$ and the normalized central Schmidt rank $\chicut(l)/\chimax$.}
  Upper row: $\IF(l)/N$. Lower row: $\chicut(l)/\chimax$. Columns correspond to $h_z=0,0.01,0.5,1$ from left to right. Dashed curves in the lower row are logistic fits of Eq.~\eqref{eq:logistic}. As $h_z$ increases, the final infidelity decreases and the Schmidt-rank growth becomes slower.
  }
  \label{fig:encoding_IF_chi}
\end{figure*}
We obtain an approximate ground state $\psidmrg$ by DMRG with the bond dimension truncated to $\chi=N$, and use it as the classical input to the quantum-circuit encoder.
Here, $\chi=N$ is an intentionally coarse truncation that keeps the classical cost polynomial in $N$; the high-entanglement components left uncaptured at this bond dimension are recovered by the subsequent PITE.
For the disentangling circuit
\begin{equation}
  V(l)=\hat U_l\cdots \hat U_1,
\end{equation}
we read off the Schmidt rank $\chicut(l)$ across the central bond of $V(l)\psidmrg$ and analyze the ratio $\chicut(l)/\chimax$.
For an $N$-site spin-$1/2$ chain, the maximum possible Schmidt rank across the central bond is
\begin{equation}
  \chimax
  =
  2^{\lfloor N/2\rfloor},
  \label{eq:chi_max}
\end{equation}
where the central bond is understood as the cut separating $\lfloor N/2\rfloor$ sites from the remaining $N-\lfloor N/2\rfloor$ sites.
For even $N$, this reduces to $\chimax=2^{N/2}$.
As an accuracy diagnostic we use the fidelity between the encoded state $\psienc(l)$ and the exact ground state $\psiexact$ obtained by Lanczos diagonalization,
\begin{equation}
  F(l)=|\braket{\psi_{\mathrm{exact}}}{\psi_{\mathrm{enc}}(l)}|^2,
\end{equation}
from which we evaluate the per-site infidelity $\IF(l)/N$.

Figure~\ref{fig:encoding_IF_chi} shows the layer dependence of $\IF(l)/N$ and $\chicut(l)/\chimax$. In the upper row, $\IF(l)/N$ drops rapidly during the first few layers and then improves only marginally with further layers. In the lower row, $\chicut(l)/\chimax$ traces an S-shaped curve that is well fitted by the logistic function
\begin{equation}
  \frac{\chicut(l)}{\chimax}
  =
  \frac{1}{1+\left(\frac{1}{r_0}-1\right)e^{-\gamma l}} ,
  \label{eq:logistic}
\end{equation}
with $r_0=\chi_0/\chimax$ the initial normalized Schmidt rank and $\gamma$ the growth rate. The inflection point of this fit,
\begin{equation}
  \Lstar
  =
  \frac{1}{\gamma}
  \ln\left(\frac{1}{r_0}-1\right) ,
  \label{eq:lstar}
\end{equation}
corresponds to $\chicut(\Lstar)=\chimax/2$.

As Fig.~\ref{fig:encoding_IF_chi} shows, this $\Lstar$ is the inflection point of $\chicut(l)/\chimax$ and, at the same time, approximately coincides with the layer at which $\IF(l)/N$ in the upper row crosses over from a rapid decrease to a slow improvement. The single scale $\Lstar$ thus characterizes the turning point of both the entanglement growth of the intermediate state and the fidelity improvement.

\subsection{\label{sec:rank_growth_origin}Origin of rank growth and scrambling-like interpretation}

The observed growth of $\chicut(l)$ with $l$ is counterintuitive, since the disentangling circuit is designed to push the target state toward a product state. This growth can be attributed to the structure of the Ran-type MPD encoder, in which each layer is constructed from a rank-2 truncated MPS rather than from the full intermediate state~\cite{ran2020}. Let $\ket{\psi_l}$ denote the state to be disentangled at layer $l$ and $\ket{\tilde\psi_l}$ its rank-2 approximation. The MPD layer $\hat U_l$ acts mainly to disentangle the low-rank subspace spanned by $\ket{\tilde\psi_l}$, while the residual component of $\ket{\psi_l}$ orthogonal to this subspace is not removed but undergoes a nontrivial unitary rotation, and is carried into the next layer. The subsequent nearest-neighbor two-qubit gates then redistribute this residual over a wider set of degrees of freedom.

Through this mechanism, while the approximation error of the encoded low-rank component decreases, the Schmidt rank needed to faithfully represent the intermediate state can increase. As a rough estimate, each MPD layer enlarges the effective virtual dimension by a factor of the physical dimension $d$, so after $k$ layers
\begin{equation}
  \chi_{\rm eff}(k)
  \sim
  \min\left(\chi_0 d^k,\,\chimax\right) ,
\end{equation}
where $\chi_0$ is the initial effective Schmidt rank and $\chimax=2^{N/2}$ is the maximum Schmidt rank allowed at the central bond. Initially, the rank growth is approximately exponential, and saturates as the finite Hilbert-space bound is approached. The logistic curve in Eq.~\eqref{eq:logistic} can thus be understood as a minimal effective model for this competition between residual proliferation and finite-size saturation. The growth rate $\gamma$ is an effective parameter that captures not only the bare circuit depth but also the size of the residual discarded by the rank-2 truncation, the extent to which the residual spreads under subsequent MPD layers, and the correlation structure of the target state.

The redistribution of this residual information is analogous to scrambling in quantum circuits, in which initially local information or operators spread over many degrees of freedom of a many-body system and become difficult to recover from local observables alone. In random brick-wall circuits, operator fronts propagate at a butterfly velocity, and the saturation time in finite-size systems is known to scale linearly with system size~\cite{nahum2018}. The Ran-type sequential MPD circuit is not a random circuit and its gates are arranged sequentially, so the quantitative butterfly velocity or out-of-time-order correlator (OTOC) results from random-circuit theory do not directly apply. Nevertheless, the phenomenology remains consistent with that of scrambling in finite-size quantum circuits. (i) The residual arises from local truncation errors. (ii) Subsequent gates spread it over a wider set of degrees of freedom. (iii) The Schmidt rank crosses over from early exponential growth to finite-size saturation. (iv) The inflection point scales approximately as $\Lstar=\order{N}$. On this basis, we interpret the logistic growth of $\chicut(l)$ as a scrambling-like spreading of information internal to the sequential MPD encoder.

\subsection{\label{sec:gamma_gap}Dependence of $\gamma$ on gap and system size}

\begin{figure}[t]
  \centering
  \includegraphics[width=\linewidth]{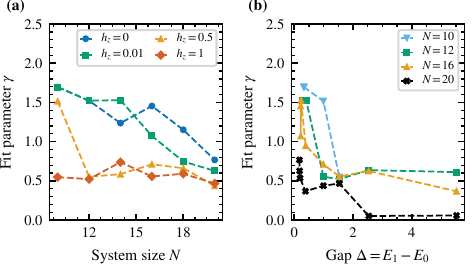}
  \caption[Growth-rate scaling]{
  \textbf{System-size and gap dependence of the logistic growth rate $\gamma$.}
  Markers show the fitted growth rates for each $N$ and $h_z$. A larger gap localizes correlations and suppresses the long-range spreading of the residual, lowering $\gamma$.
  }
  \label{fig:gamma_scaling}
\end{figure}

Figure~\ref{fig:gamma_scaling} shows the dependence of the fitted growth rate $\gamma$ on the gap and the system size.
Over the accessible range $\gamma$ tends to decrease as the energy gap $\Delta=E_1-E_0$ grows. This trend is consistent with the exponential clustering theorem for gapped local Hamiltonians~\cite{nachtergaele2006}. For any local operators $A\in\mathcal{A}_x$ and $B\in\mathcal{A}_Y$,

\begin{equation}
 \left|
 \langle\psi_0|AB|\psi_0\rangle
 -
 \langle\psi_0|A|\psi_0\rangle
 \langle\psi_0|B|\psi_0\rangle
 \right|
 \le
 c(A,B)\,e^{-\mu d(x,Y)} ,
 \label{eq:clustering}
\end{equation}
with the correlation length $\xi=1/\mu$ satisfying $\xi\sim\Delta^{-1}$ in the small-gap limit. A larger $\Delta$ thus localizes the quantum correlations and reduces the long-range residual lost by the rank-2 truncation. The effective spreading of that residual under subsequent MPD layers is suppressed, and $\gamma$ decreases.

In the critical case $h_z=0$, correlations decay only algebraically and the central entanglement entropy grows as $S(N/2)\sim(c/6)\log N$~\cite{calabrese2004}. The rank-2 truncation error therefore grows with system size, and $\gamma$ shows a clearer $N$ dependence. In the gapped case, the entanglement saturates with $N$ owing to the area law~\cite{hastings2007}, and the $N$ dependence of $\gamma$ is correspondingly suppressed.

\subsection{\label{sec:lstar_accuracy}Scaling of $\Lstar$ and attainable accuracy}

\begin{figure}[t]
  \centering
  \includegraphics[width=\linewidth]{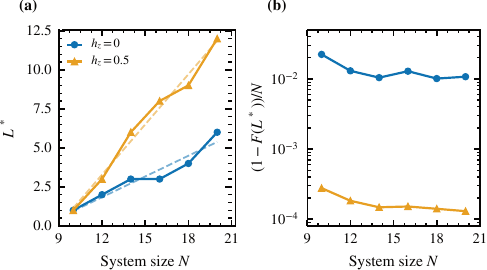}
  \caption[Efficient encoding boundary]{
  \textbf{Efficient encoding boundary $\Lstar$ and the accuracy reached there.}
  (a) System-size dependence of the inflection point $\Lstar$, with dashed linear fits.
  (b) Per-site infidelity $\IF(\Lstar)/N$ at the inflection point. Within this size range $\Lstar$ scales approximately linearly with $N$, while $\IF(\Lstar)/N$ saturates with $N$.
  }
  \label{fig:lstar_accuracy}
\end{figure}

Figure~\ref{fig:lstar_accuracy} shows the system-size scaling of the inflection point $\Lstar$ and of the accuracy reached there, $\IF(\Lstar)/N$. By definition $\Lstar$ is the layer at which $\chicut/\chimax=1/2$, so it tracks the exponential growth of $\chimax=2^{N/2}$. Over the accessible range, both $h_z=0$ and $h_z=0.5$ follow
\begin{equation}
  \Lstar = \order{N} .
  \label{eq:lstar_linear}
\end{equation}

The accuracy diagnostic $\IF(\Lstar)/N$ saturates with $N$. This indicates that sequential MPS encoding up to $\Lstar$ efficiently captures the short-range correlations within the correlation length, while leaving behind the high-entanglement components that the rank-2 truncation cannot represent. The saturation value depends strongly on $h_z$, being larger in the critical case and smaller in the gapped case. The $\Lstar$-encoded state is therefore not yet a ground state at the target accuracy, but functions as a high-quality initial state for the subsequent PITE refinement.

\subsection{\label{sec:tail_barrier}Encoding barrier beyond $\Lstar$}

\begin{figure}[t]
  \centering
  \includegraphics[width=\linewidth]{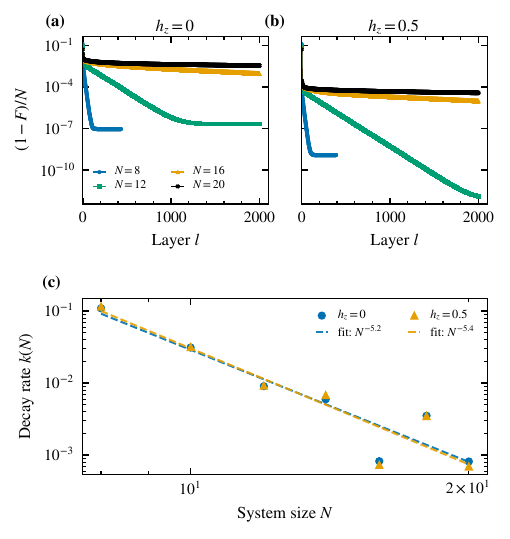}
  \caption[Encoding barrier beyond Lstar]{
  \textbf{Encoding barrier beyond $\Lstar$.}
  In the tail region, $\IF(l)/N$ decays exponentially with $l$, but the fitted decay rate $k(N)$ empirically shrinks as $N^{-5}$. An encoding-only strategy aimed at arbitrary accuracy therefore scales as $\order{N^5\log(N/\eps)}$.
  }
  \label{fig:encoding_barrier}
\end{figure}

Beyond $\Lstar$, $\IF(l)/N$ keeps decreasing, but the rate of improvement drops sharply. We fit the semilog-linear tail by
\begin{equation}
  \frac{\IF(l)}{N}
  =
  C(N)+A(N)e^{-k(N)l}
  \label{eq:tail_fit}
\end{equation}
and extract the decay rate $k(N)$. As shown in Fig.~\ref{fig:encoding_barrier}, both $h_z=0$ and $h_z=0.5$ follow the empirical scaling
\begin{equation}
  k(N)\propto N^{\alpha} ,
  \qquad
  \alpha\simeq -5 .
  \label{eq:k_scaling}
\end{equation}

Combining Eqs.~\eqref{eq:tail_fit} and \eqref{eq:k_scaling}, the number of additional layers needed beyond $\Lstar$ to reach a target infidelity $\eps$ is
\begin{equation}
  \Delta L
  \sim
  \frac{1}{k(N)}
  \log\frac{A(N)}{\eps/N-C(N)}
  \propto
  N^5\log\frac{N}{\eps} .
  \label{eq:deltaL}
\end{equation}
Pushing the tail by deeper encoding alone therefore costs a quantum-circuit depth of order $\order{N^5\log(N/\eps)}$, much greater than the $\order{N}$ depth that brings the encoder to $\Lstar$. The encoding-only route thus becomes inefficient rapidly with increasing $N$, and refining the state beyond $\Lstar$ by a mechanism other than deeper encoding is a natural alternative.

\section{\label{sec:pite}PITE Refinement and Circuit-Level Implementation}
% ============================================================

\subsection{\label{sec:pite_initializer}$\Lstar$-MPS encoded state as a PITE initializer}

Section~\ref{sec:encoding} showed that sequential MPS encoding transfers the ground-state component to the quantum register efficiently up to $\Lstar$, but the rate of improvement drops sharply in the tail beyond $\Lstar$. We therefore stop the encoder at $\Lstar$ rather than push it to arbitrary accuracy, and use the resulting state as the PITE initial state. From the DMRG MPS we construct the $\Lstar$-layer encoding circuit and define
\begin{equation}
  \ket{\psi_{\rm MPS}}
  \equiv
  \psienc(\Lstar)
  =
  U_{\rm enc}(\Lstar)\ket{0}^{\otimes N} .
  \label{eq:psi_mps_init}
\end{equation}
For comparison we use the simple antiferromagnetic product state
\begin{equation}
  \ket{\psi_{\rm Néel}}
  =
  \ket{0101\cdots} ,
  \label{eq:neel_init}
\end{equation}
which reflects the classical order of the model but carries no quantum correlations. In contrast, the $\Lstar$-MPS encoded state already encodes the short- and intermediate-range correlations of the DMRG ground state into the quantum circuit. This difference appears not only in the post-PITE energy error but, most notably, in the cumulative success probability $P_{\rm cum}$.

\subsection{\label{sec:deterministic_schedule}DMRG-informed deterministic linear schedule}

We fix the PITE imaginary-time-evolution schedule deterministically from the classical information provided by a coarse, low-bond-dimension DMRG calculation. The required inputs are the DMRG ground-state energy $E_0^{\rm DMRG}$, the effective finite-size gap $\Delta_{\rm eff}^{\rm DMRG}$, and the reference ground-state weight of the initial state,
\begin{equation}
  w_0^{\rm ref}
  =
  \left|\braket{\psi_{\rm DMRG}}{\psi_{\rm init}}\right|^2 ,
  \label{eq:w0_ref}
\end{equation}
where $\ket{\psi_{\rm init}}$ is either the state in Eq.~\eqref{eq:psi_mps_init} or that in Eq.~\eqref{eq:neel_init}. For numerical validation we evaluate infidelities and energy errors against the exact ground state obtained by Lanczos diagonalization.

We use the DMRG energy as the energy shift,
\begin{equation}
  E_{\rm shift}=E_0^{\rm DMRG} ,
  \label{eq:eshift_dmrg}
\end{equation}
so that the PITE filter has maximum success amplitude on the ground-state component. For an excited state $i$ with excitation energy $\Delta_i=E_i-E_0$, the cumulative filter factor reads approximately
\begin{equation}
  \prod_{k=1}^{K}
  \cos^2\!\left(\Delta_i s\dtau_k\right) .
  \label{eq:pite_filter_product}
\end{equation}
The largest imaginary-time step is set so that the lowest relevant excitation is sufficiently suppressed:
\begin{equation}
  s\dtau_{\max}
  =
  \frac{0.62\pi}{\Delta_{\rm eff}^{\rm DMRG}} .
  \label{eq:dtau_max_gap}
\end{equation}
The factor $0.62\pi$ is the location of the minimum of the single-eigenvalue filter in the linear-scheduling analysis of Nishi \textit{et al.}~\cite{nishi2023optimal}. The minimum step $s\dtau_{\min}$ is fixed at a small value, and the intermediate steps are distributed linearly,
\begin{equation}
\begin{aligned}
  \dtau_k
  =&\, \dtau_{\min}
  +
  \frac{k-1}{K_{\rm safe}-1}
  \left(\dtau_{\max}-\dtau_{\min}\right) , \\
  & k=1,\ldots,K_{\rm safe} .
\end{aligned}
  \label{eq:linear_schedule_safe}
\end{equation}

The total number of PITE steps is fixed by the conservative linear-scheduling estimate of Nishi \textit{et al.},
\begin{equation}
  K_{\rm safe}
  =
  \left\lceil
  \frac{3}{2\ln 2}
  \ln
  \left[
  \frac{1-w_0^{\rm ref}}
       {\tilde{\eps}\,w_0^{\rm ref}}
  \right]
  \right\rceil ,
  \qquad
  \tilde{\eps}
  =
  \frac{\eps(4-\eps)}{(2-\eps)^2} .
  \label{eq:K_safe}
\end{equation}
Equation~\eqref{eq:K_safe} is a sufficient-condition estimate obtained from the interval of the linear schedule where the $\cos^2$ factor is small, and is more conservative than the average-large-eigenvalue estimate~\cite{nishi2023optimal}. The PITE trajectory in the present work is fixed in advance by Eqs.~\eqref{eq:eshift_dmrg}--\eqref{eq:K_safe}.

\subsection{\label{sec:pite_circuit}RZZ-native second-order Trotter implementation}

We implement the real-time-evolution operator $\hat U_{\rm RTE}$ inside the PITE circuit with a second-order Trotter formula. We split the Hamiltonian as
\begin{equation}
  H
  =
  H_{\rm even}+H_{\rm odd}+H_{\rm field} ,
  \label{eq:H_even_odd_field}
\end{equation}
where $H_{\rm even}$ and $H_{\rm odd}$ are the Heisenberg interactions on even and odd bonds and $H_{\rm field}$ is the staggered-field term. One Trotter step is then the symmetric decomposition~\cite{Trotter1959,Suzuki1976,suzuki1990,childs2021trotter}
\begin{equation}
\begin{aligned}
  S_2(\delta t)
  =&\, e^{-i(\delta t/2)H_{\rm even}}
  e^{-i(\delta t/2)H_{\rm odd}}
  e^{-i\delta t H_{\rm field}} \\
  &\times
  e^{-i(\delta t/2)H_{\rm odd}}
  e^{-i(\delta t/2)H_{\rm even}} .
\end{aligned}
  \label{eq:strang_eof}
\end{equation}

At the circuit level we take $R_{ZZ}$, $R_Z$, and $R_X$ as native gates. With the convention
\begin{equation}
  R_{ZZ}(\theta)=\exp\left[-i\frac{\theta}{2}Z\otimes Z\right] ,
  \qquad
  R_Z(\phi)=\exp\left[-i\frac{\phi}{2}Z\right] ,
  \label{eq:rzz_convention}
\end{equation}
the $XX$ and $YY$ bond terms are mapped to $R_{ZZ}$ by single-qubit basis changes. The forward and backward controlled real-time evolutions in PITE are realized as $R_{ZZ}$ gates acting between the ancilla and the system, with effective interaction of the form $Z_a\otimes H$. Appendix Fig.~\ref{fig:urte} details the construction. This route exploits the native $R_{ZZ}$ structure directly and avoids the explicit decomposition of controlled-$U$ and controlled-$U^\dagger$ gates.

The PITE schedule of Eqs.~\eqref{eq:eshift_dmrg}--\eqref{eq:K_safe} is fixed before the circuit simulation. In the present numerical validation, the Trotter repetition number for each PITE step is calibrated by state-vector simulation to evaluate the circuit depth, and we take the smallest repetition number that satisfies the assigned error budget. The same calibration rule applies to both the MPS-encoded and N\'eel initializers. Details of the numerical PITE protocol and of how target-reaching resources are extracted are summarized in Appendix~\ref{app:pite_protocol}. All circuits are constructed in Qiskit~\cite{qiskit2024}, transpiled at optimization level 3 with $R_{ZZ}$, $R_Z$, and $R_X$ as native gates, and the circuit depth is measured on the transpiled circuit.

\subsection{\label{sec:pite_numerics}Numerical setup and representative $N=16$ benchmark}

The numerical study uses Eq.~\eqref{eq:ham_spin} at $h_z=0$ and $h_z=0.5$: the former is a gapless chain with a vanishing finite-size gap, and the latter a gapped chain due to the staggered field. We set the maximum DMRG bond dimension to $\chi=N$ and run the sequential MPD encoder up to $\Lstar$ as defined in Sec.~\ref{sec:lstar_accuracy}. The two initializers we compare are $\ket{\psi_{\rm MPS}}=\psienc(\Lstar)$ and $\ket{\psi_{\rm N\acute eel}}$, and we run both through the same deterministic schedule, the same quantum-circuit implementation, and the same Trotter calibration rule. At the point of reaching chemical accuracy ($\Delta E_{\rm chem}=1.5936\times10^{-3}J$), we record the raw depth, the success-probability-weighted depth, and the cumulative success probability.

\begin{figure}[t]
  \centering
  \includegraphics[width=\linewidth]{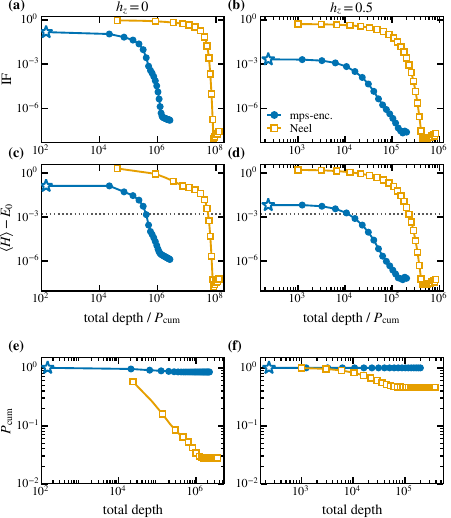}
  \caption[PITE trajectories for N=16]{
  \textbf{PITE refinement trajectories for $N=16$.}
  Left and right columns: $h_z=0$ and $h_z=0.5$. Upper, middle, and lower panels: infidelity, energy error, and cumulative success probability. Blue circles, $\Lstar$-MPS encoded initial state; orange squares, N\'eel initial state. The MPS-encoded initializer substantially improves the cumulative success probability in particular, and lowers the post-selection-weighted effective cost.
  }
  \label{fig:n16_trajectories}
\end{figure}

Figure~\ref{fig:n16_trajectories} shows representative PITE refinement trajectories for $N=16$. The upper, middle, and lower panels show the infidelity, the energy error, and the cumulative success probability, respectively. The horizontal axis is the post-selection-weighted effective depth in the upper and middle panels and the raw depth in the lower panels. For both $h_z=0$ and $h_z=0.5$, the MPS-encoded initializer reaches the target region at a shallower effective depth than the N\'eel initializer. The cumulative success probability separates the two initializers even more clearly. For the MPS-encoded initializer, $P_{\rm cum}$ stays of order unity, whereas for the N\'eel initializer it drops substantially, especially in the gapless chain. Since the expected number of trials for a successful PITE trajectory is $\propto 1/P_{\rm cum}$, this difference translates directly into a reduction of the post-selection overhead.

% ============================================================
\section{\label{sec:resource}Circuit Resource Scaling}
% ============================================================

\subsection{\label{sec:resource_measures}Resource measures}

We now analyze the circuit resources required for PITE refinement. Three measures are used throughout. The raw PITE depth for a single successful trajectory is
\begin{equation}
  D_{\rm raw}
  =
  \sum_{k=1}^{K_{\rm chem}} d_k ,
  \label{eq:D_raw}
\end{equation}
with $d_k$ the transpiled depth of the $k$th PITE step and $K_{\rm chem}$ the number of steps needed for the energy error to fall below $\Delta E_{\rm chem}$. The effective depth that absorbs the post-selection overhead is
\begin{equation}
  D_{\rm post}
  =
  \frac{D_{\rm raw}}{P_{\rm cum}} .
  \label{eq:D_post}
\end{equation}
The MPS-encoding cost up to $\Lstar$ is treated separately. Since the sequential MPD encoder has $\Lstar=\order{N}$,
\begin{equation}
  D_{\rm enc}(\Lstar)=\order{N} .
  \label{eq:G_enc_Lstar}
\end{equation}
By contrast, the encoding-only tail strategy of Sec.~\ref{sec:tail_barrier} requires
\begin{equation}
  D_{\rm tail}=\order{N^5\log\frac{N}{\eps}} .
  \label{eq:G_tail_again}
\end{equation}
The quantities $D_{\rm raw}$ and $D_{\rm post}$ compared below are thus the PITE-refinement costs that follow stopping the encoder at $\Lstar$.

\subsection{\label{sec:analytical_depth_scaling}Analytical depth estimate}

The leading $N$ dependence of the PITE refinement depth comes from the controlled real-time evolution used in each PITE step.
We consider the direct Trotterized implementation of the real-time evolution generated by the local Hamiltonian
\begin{equation}
  H=\sum_{j\in\mathcal{J}} h_j ,
\end{equation}
where $\mathcal{J}$ denotes the set of local Pauli terms in Eq.~\eqref{eq:ham_pauli}.
For the staggered-field Heisenberg chain, $|\mathcal{J}|=\order{N}$ and each local term has bounded operator norm.

For the second-order Trotter--Suzuki formula, the relevant commutator prefactor can be written as
\begin{equation}
  \alpha_{\rm comm}
  =
  \sum_{a,b,c\in\mathcal{J}}
  \left\|
    \left[
      h_c,
      \left[
        h_b,h_a
      \right]
    \right]
  \right\| .
  \label{eq:alpha_comm_definition}
\end{equation}
This is the $p=2$ specialization of the commutator-scaling prefactor used in the general $p$th-order estimate of Ref.~\cite{nishi2023optimal}.
For a one-dimensional nearest-neighbor Hamiltonian, a nested commutator in Eq.~\eqref{eq:alpha_comm_definition} is nonzero only when the corresponding local terms have overlapping or adjacent supports.
Each local term therefore contributes to only $\order{1}$ nonzero nested commutators, and each such commutator has $\order{1}$ norm.
Since the number of local terms is $\order{N}$, we obtain
\begin{equation}
  \alpha_{\rm comm}=\order{N}.
  \label{eq:comm_extensive}
\end{equation}

Let $t=s\dtau$ be the real-time evolution length appearing in one PITE step, and let $\eta_{\rm TS}$ denote the Trotter-error budget assigned to that real-time evolution block.
The commutator-scaling estimate for the second-order formula gives
\begin{equation}
  r(t)
  =
  \order{
    \alpha_{\rm comm}^{1/2}
    t^{3/2}
    \eta_{\rm TS}^{-1/2}
  } .
  \label{eq:trotter_second_order}
\end{equation}
With a fixed Trotter-error budget and $\alpha_{\rm comm}=\order{N}$, this becomes
\begin{equation}
  r(t)=\order{N^{1/2}t^{3/2}} .
  \label{eq:r_scaling}
\end{equation}

In the direct controlled-RTE implementation used in this work, one Trotter repetition contains $\order{N}$ controlled local rotations, including the nearest-neighbor interaction terms and the field terms.
With the transpiled-depth convention used in our circuit-level resource estimate, the depth per Trotter repetition is therefore $\order{N}$.
The depth of one PITE step with real-time scale $t$ is then estimated as
\begin{equation}
  d_{\rm step}(t)
  =
  \order{N\,r(t)}
  =
  \order{N^{3/2}t^{3/2}} .
  \label{eq:d_step_scaling}
\end{equation}

For a schedule consisting of $K$ PITE steps, the raw PITE depth satisfies
\begin{equation}
  D_{\rm raw}
  =
  \sum_{k=1}^{K} d_{\rm step}(t_k)
  \le
  K\,d_{\rm step}(t_{\max}) .
  \label{eq:D_raw_sum_bound}
\end{equation}
The number of PITE steps contributes logarithmic factors in the target accuracy and the initial ground-state weight, as in the scheduling analysis of Ref.~\cite{nishi2023optimal}.
Here we focus on the leading dependence on $N$ and on the effective gap, and suppress these logarithmic and overlap-dependent factors.

The maximum real-time scale of the linear schedule is set by the inverse of the effective gap,
\begin{equation}
  t_{\max}=s\dtau_{\max}
  \sim
  \Delta_{\rm eff,N}^{-1} .
  \label{eq:tmax_gap}
\end{equation}
Substituting this scale into Eq.~\eqref{eq:d_step_scaling} gives the leading gap-controlled estimate
\begin{equation}
  D_{\rm raw}
  =
  \order{
    N^{3/2}\Delta_{\rm eff,N}^{-3/2}
  } ,
  \label{eq:D_raw_gap_scaling}
\end{equation}
up to the suppressed scheduling and post-selection factors.

In the gapless chain at $h_z=0$, the finite-size gap closes as
\begin{equation}
  \Delta_{\rm eff,N}\sim N^{-1} ,
  \label{eq:gapless_gap}
\end{equation}
which combined with Eq.~\eqref{eq:D_raw_gap_scaling} gives
\begin{equation}
  D_{\rm raw}(h_z=0)
  =
  \order{N^3} .
  \label{eq:D_gapless}
\end{equation}
In the gapped chain at $h_z=0.5$,
\begin{equation}
  \Delta_{\rm eff,N}\to\Delta_\infty>0 ,
  \label{eq:gapped_gap}
\end{equation}
so
\begin{equation}
  D_{\rm raw}(h_z=0.5)
  =
  \order{N^{3/2}} .
  \label{eq:D_gapped}
\end{equation}
We test this analytical gap-controlled picture against circuit-level numerical data below.

\subsection{\label{sec:numerical_depth_scaling}Numerical scaling under the deterministic schedule}

\begin{figure}[t]
  \centering
  \includegraphics[width=\linewidth]{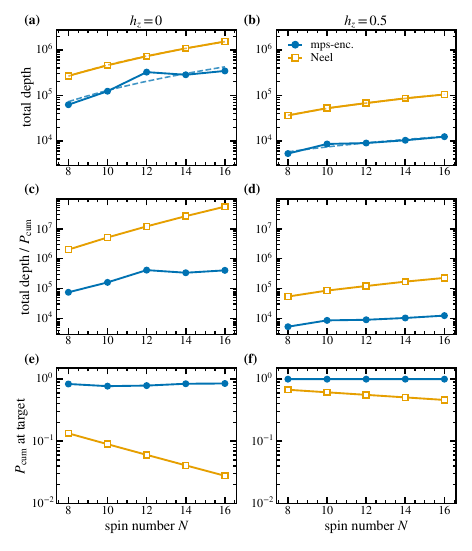}
  \caption{
  \textbf{System-size dependence of the PITE refinement cost.}
  Left column: $h_z=0$. Right column: $h_z=0.5$. Upper, middle, and lower panels: raw depth at chemical accuracy, $D_{\rm raw}/P_{\rm cum}$, and cumulative success probability at the target. Blue circles, $\Lstar$-MPS encoded initial state; orange squares, N\'eel initial state. The MPS-encoded initializer, particularly by keeping $P_{\rm cum}=O(1)$, substantially reduces the post-selection-weighted effective cost.
  }
  \label{fig:pite_scaling}
\end{figure}

Figure~\ref{fig:pite_scaling} shows the system-size dependence of the PITE refinement cost for $N=8,10,12,14,16$. The upper, middle, and lower panels show the raw depth at chemical accuracy, the post-selection-weighted depth $D_{\rm post}=D_{\rm raw}/P_{\rm cum}$, and the cumulative success probability at that target. All data points are obtained under the same deterministic schedule defined by Eqs.~\eqref{eq:eshift_dmrg}--\eqref{eq:K_safe}.

Table~\ref{tab:depth_fit} lists the log--log fits of the raw depth to $D_{\rm raw}=\alpha N^\beta$. The N\'eel initial state in the gapped chain at $h_z=0.5$ gives
\begin{equation}
  \beta=1.521\pm0.023 ,
\end{equation}
in good agreement with the $N^{3/2}$ prediction of Eq.~\eqref{eq:D_gapped}. The MPS-encoded initial state in the same gapped chain yields a smaller exponent, $\beta = 1.117 \pm 0.188$. This may reflect the effective commutator prefactor obtained through the state-dependent Trotter calibration may scale more weakly than the worst-case $\alpha_{\rm comm}=O(N)$ estimate, which can further reduce the observed exponent.

For the gapless chain at $h_z=0$, both the N\'eel and MPS-encoded initial states give finite-size effective exponents around $\beta\simeq2.5$. Importantly, the growth in the gapless chain is clearly steeper than in the gapped chain, in line with the analytical picture of Eq.~\eqref{eq:D_raw_gap_scaling} that includes the additional factor $\Delta_{\rm eff,N}^{-3/2}\sim N^{3/2}$ from the closing finite-size gap.

At the same time, the observed exponent is smaller than the conservative $\order{N^3}$ worst-case estimate. This deviation can be attributed to the calibration of the Trotter repetition numbers in the numerical simulations. They are calibrated state-dependently along the actual PITE trajectory, rather than obtained from a worst-case operator-norm bound over the full Hilbert space. If the worst-case commutator prefactor is replaced by an effective one, $\alpha_{\rm eff}=\order{N^{a_{\rm eff}}}$, then for a fixed Trotter-error budget
\begin{equation}
  D_{\rm raw}
  =
  \order{
    N\alpha_{\rm eff}^{1/2}\Delta_{\rm eff,N}^{-3/2}
  } ,
\end{equation}
where $\alpha_{\rm eff}$ is distinct from the fit prefactor $\alpha$ in Table~\ref{tab:depth_fit}. Since $\Delta_{\rm eff,N}\sim N^{-1}$ in the gapless chain, this gives
\begin{equation}
  D_{\rm raw}
  =
  \order{
    N^{5/2+a_{\rm eff}/2}
  } .
\end{equation}
The observed $\beta\simeq2.5$ therefore suggests that, in the present finite-size range, the effective commutator prefactor sampled by the state-dependent calibration behaves closer to $\order{1}$ than to the conservative $\order{N}$ worst-case scaling.

\begin{table}[t]
  \centering
  \caption{
  Power-law fits $D_{\rm raw}=\alpha N^\beta$ for the raw depth at chemical accuracy, performed in log--log scale over $N=8,10,12,14,16$. The exponents should be read as finite-size effective exponents.
  }
  \label{tab:depth_fit}
  \begin{adjustbox}{max width=\linewidth}
  \begin{tabular}{cccccc}
    \toprule
    $h_z$ & initial state & $\alpha$ & $\beta$ & standard error & $R^2_{\log}$ \\
    \midrule
    $0$   & $\Lstar$-MPS encoded & $3.674\times10^2$ & $2.543$ & $0.566$ & $0.871$ \\
    $0$   & N\'eel               & $1.432\times10^3$ & $2.510$ & $0.026$ & $1.000$ \\
    $0.5$ & $\Lstar$-MPS encoded & $5.632\times10^2$ & $1.117$ & $0.188$ & $0.922$ \\
    $0.5$ & N\'eel               & $1.549\times10^3$ & $1.521$ & $0.023$ & $0.999$ \\
    \bottomrule
  \end{tabular}
  \end{adjustbox}
\end{table}

\subsection{\label{sec:success_overhead}Post-selection overhead and comparison with encoding-only strategy}

The leading $N$ dependence of the raw depth is set by the gap and the Trotter cost, but for the effective resource cost of PITE the cumulative success probability matters just as much. Figures~\ref{fig:pite_scaling}(e) and \ref{fig:pite_scaling}(f) show that the MPS-encoded initializer keeps $P_{\rm cum}=\order{1}$ for both $h_z=0$ and $h_z=0.5$. The N\'eel initializer, in contrast, loses $P_{\rm cum}$ rapidly with system size, especially in the gapless chain. The advantage of the MPS-encoded initializer is therefore even more visible in $D_{\rm raw}/P_{\rm cum}$ shown in the middle panels.

Combined with the encoding barrier of Sec.~\ref{sec:tail_barrier}, we obtain the following picture. Pushing beyond $\Lstar$ by adding more MPD layers requires a tail depth of $\order{N^5\log(N/\eps)}$. When the $\Lstar$-encoded state is used as the PITE initial state instead, imaginary-time evolution suppresses the residual, and even in the gapless chain the fidelity is recovered at a lower cost than this tail strategy. Furthermore, this initial state carries a large $|c_0|^2$ and a high $P_{\rm cum}$, and thus substantially reduces the post-selection overhead. Stopping the encoder at $\Lstar$ and refining the state with a deterministically scheduled PITE protocol therefore offers a practical resource balance.

% ============================================================
\section{\label{sec:discussion}Discussion}
% ============================================================

The scalings of $\Lstar$ and $k(N)$ obtained here reveal an intrinsic limit of optimization-free encoders built from sequential MPD layers and, at the same time, provide a practical stopping rule for their use.
The weak dependence of these scalings on the presence or absence of a spectral gap suggests that the tail-region inefficiency is not governed solely by the correlation length of the target Hamiltonian.
Rather, it also reflects structural features of the encoder itself: the rank-$2$ truncation used to construct each MPD layer, the propagation of the residual component outside the truncated subspace, and the directional asymmetry between the construction of the disentangling layers and the execution of the corresponding encoding circuit.

The depth scalings obtained in this work are summarized in Table~\ref{tab:depth_scaling_summary}, where $\eps$ denotes the prescribed accuracy for the global infidelity $\IF=1-F$.
The encoding-related quantities $D_{\rm enc}$ and $D_{\rm tail}$ refer to the depth of the sequential MPD circuit, while $D_{\rm raw}$ and $D_{\rm post}$ refer to the transpiled depth of the PITE circuit before and after post-selection weighting.

\begin{table}[ht]
  \centering
  \caption{
  Summary of the circuit-depth scalings obtained in this work. A second-order Trotter decomposition is used for evaluating the PITE circuit cost.
  }
  \label{tab:depth_scaling_summary}
  \begin{adjustbox}{max width=\textwidth}
  \begin{tabular}{ll}
    \toprule
    target &  scaling  \\
    \midrule
    MPD circuit up to $\Lstar$
      & $\order{N}$ \\
    MPD circuit beyond $\Lstar$
      & $\order{N^5\log(N/\eps)}$ \\
    PITE for a gapless system ($h_z=0$)
      & $\order{N^3}$ \\
    PITE for a gapped system ($h_z=0.5$)
      & $\order{N^{3/2}}$ \\
    \bottomrule
  \end{tabular}
  \end{adjustbox}
\end{table}

These estimates should be distinguished from black-box query complexities for spectral filtering algorithms.
Near-optimal ground-state preparation based on block-encoding and quantum signal processing provides powerful oracle-level guarantees when one is given a Hamiltonian block-encoding $U_H$, an initial-state preparation oracle $U_I$ with nontrivial ground-state overlap $\gamma$, and a lower bound on the relevant spectral gap~\cite{lintong2020}.
Such results are fundamental benchmarks in the black-box model, but the required oracles themselves are not supplied by the theory.
In particular, the construction of a high-overlap initial state and the realization of a suitable Hamiltonian block-encoding remain implementation-level tasks.

The present workflow addresses this complementary, implementation-oriented problem.
The initial state is not assumed as a black-box oracle: it is constructed explicitly from a DMRG-derived MPS by the sequential MPD encoder.
Likewise, the refinement stage does not require a Hamiltonian block-encoding or QSP phase-factor synthesis; it uses controlled real-time evolutions for the local Hamiltonian, whose Trotterized circuit depth can be estimated directly.
Thus the algorithm is specified in terms of concrete inputs and operations---DMRG tensors, an approximate ground-state energy, an effective gap, a reference overlap, an MPD circuit, and a deterministic PITE schedule---rather than in terms of abstract oracle calls.
This explicit construction is a practical advantage of the DMRG--MPD--PITE framework, especially for early fault-tolerant or near-term settings where circuit depth, post-selection probability, and Trotterization overhead are the relevant resources.

This distinction does not imply that the present method improves the generic black-box lower bounds.
Rather, it uses information that is absent from the black-box model: locality, low-entanglement tensor-network structure, and a classically optimized DMRG approximation.
These structural ingredients are precisely what allow the high-overlap initial state to be generated constructively.
The role of the MPD encoder is therefore not merely to compress a known MPS into a circuit, but to convert classical tensor-network information into a quantum initial state whose overlap is large enough to make subsequent probabilistic refinement efficient.

The importance of this explicit state-preparation step is reinforced by the cost structure of PITE.
For probabilistic imaginary-time evolution, the effective cost is governed by the ground-state weight $|c_1|^2$ in the initial state and scales as
\[
  \order{
  \frac{d_{\rm PITE}}{|c_1|^2}
  \log\!\left(\frac{1}{\eps |c_1|^2}\right)
  },
\]
up to implementation-dependent factors~\cite{nishi2023optimal}.
The DMRG--MPD encoded state therefore acts as a state-preparation preconditioner for PITE.
In the staggered-field Heisenberg chains studied here, this preconditioning keeps the cumulative success probability of order unity and improves it substantially relative to the N\'eel product state.
Consequently, the proposed stopping rule at $\Lstar$ is not only an encoding-efficiency criterion, but also a criterion for handing over the state to PITE before the MPD tail region becomes prohibitively inefficient.

Variational optimization of the per-layer unitaries, such as the recently proposed Schmidt spectrum optimization (SSO)~\cite{green2025}, can mitigate the accumulation of truncation errors.
The present proposal does not compete with such methods in the sense of seeking the most accurate MPS encoder at fixed depth.
Instead, it provides a diagnostic and resource-estimation framework for deciding when an encoding-only strategy ceases to be efficient.
Even if a more accurate encoder shifts $\Lstar$ or modifies the tail decay rate $k(N)$, an analogous stopping point and a corresponding transition to PITE refinement can in principle be defined.
From this perspective, the scrambling-limited analysis can serve as a guide for combining improved MPS encoders with nonvariational refinement.

As alternatives to the sequential MPD encoder, recent works have shown that MPS preparation can be substantially shortened through renormalization-group constructions and adaptive quantum circuits combining mid-circuit measurements with classical feedforward.
Malz \textit{et al.}\ proved that unitary-only preparation of translation-invariant normal MPS requires a circuit depth of $T=\Omega(\log N)$, gave a renormalization-group algorithm that achieves the optimal $T=\order{\log(N/\eps)}$, and showed that allowing mid-circuit measurements with classical feedforward enables an exponential speedup to $T=\order{\log\log(N/\eps)}$~\cite{malz2024}.
Smith \textit{et al.}\ further showed that a broad class of MPS can be prepared in constant depth with adaptive circuits~\cite{smith2024}.
These optimal-depth results are important theoretical benchmarks, but their assumptions differ from those considered here.
They are obtained under translation-invariance and normality assumptions, and typically treat the bond dimension $D$ as fixed independently of the system size $N$.

For gapless systems, the true ground state exhibits logarithmically growing entanglement,
\[
  S(N/2)\sim \frac{c}{6}\log N,
\]
as predicted by conformal field theory, so that the effective bond dimension required to represent the state grows polynomially as $D\sim N^{c/6}$.
Optimal-depth results derived under a fixed-$D$ assumption are therefore not expected to apply directly to such DMRG targets.
Moreover, the $\order{\log\log N}$ and constant-depth constructions rely on mid-circuit measurements and classical feedforward, which impose nontrivial implementation constraints on hardware where these capabilities are not yet stably available.
The sequential MPS encoding studied here is unitary-only, optimization-free, and applicable to general DMRG-MPS inputs that need not be translation invariant.
It therefore plays a complementary role to adaptive and optimal-depth MPS preparation methods.

On the PITE refinement side, the balance between Trotter error and hardware noise will be central in any device-level implementation.
The present work establishes a resource estimate based on ideal state vectors, explicit circuit depth, and post-selection-weighted cost.
A next step is to incorporate noise models that reflect the gate structure of the Trotterized real-time evolution, including angle-dependent noise in small-angle $R_{ZZ}$ rotations.
Such an analysis will clarify when the reduced post-selection overhead obtained from the DMRG--MPD initial state outweighs the additional circuit depth required by PITE.
Noisy simulations and hardware demonstrations are therefore natural extensions of the present framework.

% ============================================================
\section{\label{sec:conclusion}Conclusion}
% ============================================================

In summary, we have presented a three-stage workflow for ground-state preparation that combines DMRG, sequential MPS encoding, and PITE refinement.
Starting from a DMRG-generated MPS, the MPD encoder converts the classical tensor-network state into an explicit quantum circuit, while the subsequent PITE stage further suppresses excited-state components without introducing variational optimization of the quantum circuit.
For the staggered-field Heisenberg chains studied here, we found that the central-bond Schmidt rank during the encoding process is well described by logistic growth.
The inflection point $\Lstar$ of this growth serves as a natural cutoff for the efficient encoding regime.
Up to $\Lstar$, the MPD encoder prepares a DMRG-informed initial state with depth $\order{N}$ in our depth convention and with sufficient ground-state overlap for subsequent PITE refinement in the tested systems.
In the tail region beyond $\Lstar$, the fitted per-layer improvement rate decreases as $k(N)\propto N^{-5}$.
Under this empirical tail model, reducing the global infidelity to $\eps$ by further MPS encoding alone requires an additional depth of order $\order{N^5\log(N/\eps)}$, making an encoding-only strategy increasingly inefficient with system size.
These results motivate terminating the encoder at $\Lstar$ and passing the resulting state to PITE.

In our implementation, the PITE schedule is fixed deterministically from DMRG-derived quantities, including the approximate ground-state energy, the effective gap, and the reference overlap.
For the benchmark systems considered here, the MPS-encoded initial state yields a larger cumulative post-selection probability than the N\'eel product-state reference, thereby reducing the post-selection-weighted effective cost.
The resulting workflow is specified in terms of concrete classical inputs and quantum circuits, rather than an assumed high-overlap state-preparation oracle or a Hamiltonian block-encoding oracle.
In this sense, the present method provides a practical, circuit-level route for connecting classical tensor-network calculations with nonvariational quantum projection algorithms.
This should be viewed as complementary to black-box filtering approaches, rather than as an improvement of their oracle-level complexity bounds.

Although the numerical benchmarks in this work are restricted to a one-dimensional spin chain,
the operational idea can be tested in other settings where a classically optimized tensor-network
state has appreciable overlap with the target ground state.
A natural future direction is molecular electronic structure with large strongly correlated active
spaces, where DMRG-CASCI \cite{WhiteMartin1999,ChanHeadGordon2002,ChanSharma2011,WoutersVanNeck2014}
or DMRG-CASSCF \cite{ZgidNooijen2008,Ghosh2008} provides an MPS representation of an
active-space multireference wave function.
An analogous workflow would encode this active-space MPS into a quantum register and refine it
with PITE under the active-space Hamiltonian represented on that register, for example after a
fermion-to-qubit mapping \cite{SeeleyRichardLove2012}.
If the Hamiltonian is restricted to the active space, the refinement is also restricted to that
active space, while treating dynamical correlation beyond the active space would require an
enlarged orbital space or an effective Hamiltonian that incorporates those effects
\cite{Yanai2015,BaiardiReiher2020}.
Because orbital ordering, fermion-to-qubit mappings, nonlocal two-electron interactions, and the
active-space bond dimension can affect the encoding dynamics
\cite{LegezaSolyom2003,RisslerNoackWhite2006,WoutersVanNeck2014,SeeleyRichardLove2012},
the quantitative scaling must be reassessed for molecular Hamiltonians.
Extending the framework to electronic-structure problems is therefore a promising but nontrivial
direction for future work.

\begin{acknowledgments}
The authors thank Yoshinori Suga of Toyota Motor Corporation,
and Soichi Shirai, Takahiro Horiba, Yuki Sato, Nobuko Ohba,
and Seiji Kajita of Toyota Central R\&D Labs., Inc. for valuable
discussions on potential applications of the proposed framework and
for helpful comments. The authors also acknowledge the contributions and discussions provided by the members of Quemix Inc. This work is partly supported by the UTokyo Quantum Initiative. The computation was performed using the facilities of the Supercomputer Center, the Institute for Solid State Physics, the University of Tokyo (ISSPkyodo-SC-2026-Ea-0014), the TSUBAME4.0 supercomputer at the Institute of Science Tokyo, and the Supermicro ARS-111GL-DNHR-LCC and FUJITSU Server PRIMERGY CX2550 M7 (Miyabi) at Joint Center for Advanced High Performance Computing (JCAHPC).
\end{acknowledgments}

% ============================================================
% \clearpage
\appendix
% ============================================================

\section{\label{app:pite_protocol}Numerical protocol for PITE refinement}

This appendix collects the numerical conditions used for the PITE refinement in Secs.~\ref{sec:pite} and \ref{sec:resource}. As defined in Eqs.~\eqref{eq:eshift_dmrg}--\eqref{eq:K_safe}, the PITE filtering schedule is fixed in advance from the ground-state energy, the effective finite-size gap, and the initial-state reference overlap obtained from DMRG. The calculations cover $N=8,10,12,14,16$ and $h_z=0,0.5$, with the $\Lstar$-MPS encoded state and the N\'eel state as the two initializers. The DMRG output provides the ground-state energy $E_0$, the first-excited-state energy $E_1$, the finite-size gap $\Delta=E_1-E_0$, and the initial ground-state weight $F_0$. We use $E_0$ for the energy shift and as the reference for the energy error, $\Delta$ for the maximum step, and $F_0$ for $K_{\rm safe}$.

We fix the PITE circuit parameter at $m_0=0.999$, so that
\begin{equation}
  s=\frac{m_0}{\sqrt{1-m_0^2}}\simeq 22.34 .
  \label{eq:app_s_m0}
\end{equation}
The linear schedule is
\begin{equation}
\begin{split}
  \dtau_k
  &= \dtau_{\min}
  + \frac{k-1}{K_{\rm safe}-1}
  \left(\dtau_{\max}-\dtau_{\min}\right), \\
  & \qquad k=1,\ldots,K_{\rm safe} ,
\end{split}
  \label{eq:app_linear_schedule}
\end{equation}
with maximum step
\begin{equation}
  \dtau_{\max}=\frac{0.62\pi}{s\Delta} ;
  \label{eq:app_dtau_max}
\end{equation}
$\dtau_{\min}$ is a small nonzero value to avoid a zero step. We set the total infidelity target to $\eps=10^{-6}$ and split it equally between the algorithmic and Trotter budgets, $\eps_{\rm alg}=\eps_{\rm trot}=0.5\times10^{-6}$. The schedule length is taken to be
\begin{equation}
  \tilde\eps
  =
  \frac{\eps_{\rm alg}(4-\eps_{\rm alg})}{(2-\eps_{\rm alg})^2} ,
  \qquad
  K_{\rm safe}
  =
  \left\lceil
  \frac{3}{2\ln2}
  \ln\left[\frac{1-F_0}{\tilde\eps F_0}\right]
  \right\rceil ,
  \label{eq:app_Ksafe}
\end{equation}
and the PITE trajectory is run for at most $K_{\rm safe}$ steps.

The real-time evolution within each PITE step is
\begin{equation}
  U_{\rm RTE}(t_k)=\exp[-it_k(H-E_0^{\rm DMRG})] ,
  \qquad t_k=s\dtau_k ,
  \label{eq:app_urte}
\end{equation}
approximated by the second-order symmetric Suzuki--Trotter decomposition. We split the Hamiltonian into even bonds, odd bonds, and the staggered-field term, and use one Trotter slice
\begin{equation}
  S_2(\delta t_k)
  =
  E(\delta t_k/2)O(\delta t_k/2)F(\delta t_k)
  O(\delta t_k/2)E(\delta t_k/2) .
  \label{eq:app_S2}
\end{equation}
With $r_k$ repetitions,
\begin{equation}
  U_{\rm RTE}(t_k)
  \simeq
  \left[S_2(t_k/r_k)\right]^{r_k} .
  \label{eq:app_trotter_reps}
\end{equation}
The Trotter-error budget is distributed by the cubic time-dependence of the second-order Trotter error,
\begin{equation}
  \eps_{{\rm trot},k}
  =
  \eps_{\rm trot}
  \frac{\dtau_k^3}{\sum_j\dtau_j^3} .
  \label{eq:app_trotter_budget}
\end{equation}
For each step we pick $r_k$ as the smallest integer satisfying the allocated tolerance $\eps_{{\rm trot},k}$, found by comparing the exact one-step reference state with the Trotterized one-step state. The search is capped at $r_k\le1024$. This is a state-dependent numerical calibration along the actual trajectory rather than a worst-case operator-norm bound on the full Hilbert space.

For the circuit-resource evaluation we use $R_{ZZ}$, $R_Z$, and $R_X$ as the native gate set, and record the transpiled depth and the $R_{ZZ}$ count after optimization-level-3 transpilation. With $p_k$ the success probability at step $k$, the cumulative success probability is
\begin{equation}
  P_{\rm cum}(k)=\prod_{j=1}^{k}p_j .
  \label{eq:app_Pcum}
\end{equation}
Since PITE is post-selected, the main text uses $D_{\rm raw}$ together with $D_{\rm raw}/P_{\rm cum}$ as the effective cost. The target-reaching cost is obtained by linear interpolation between two adjacent PITE steps that bracket the threshold.

\begin{table}[t]
  \centering
  \caption[PITE numerical parameters]{Main parameters used in the PITE refinement calculations.}
  \label{tab:app_pite_params}
  \begin{ruledtabular}
  \begin{tabular}{ll}
  quantity & value or definition \\
  \hline
  system sizes & $N=8,10,12,14,16$ \\
  staggered fields & $h_z=0,0.5$ \\
  initial states & $\Lstar$-MPS encoded, N\'eel \\
  total IF target & $\eps=10^{-6}$ \\
  algorithmic budget & $\eps_{\rm alg}=0.5\times10^{-6}$ \\
  Trotter budget & $\eps_{\rm trot}=0.5\times10^{-6}$ \\
  PITE circuit parameter & $m_0=0.999$ \\
  maximum step & $\dtau_{\max}=0.62\pi/(s\Delta)$ \\
  step number & $K_{\rm safe}$ \\
  Trotter allocation & proportional to $\dtau_k^3$ \\
  maximum repetitions & $r_k\le1024$ \\
  native gates & $rzz$, $rx$, $rz$ \\
  % energy target & $1.5936\times10^{-3}J$ \\
  \end{tabular}
  \end{ruledtabular}
\end{table}

\section{\label{app:rzz_resource}RZZ-native resource scaling}

\begin{figure*}[ht]
  \centering
  \resizebox{\textwidth}{!}{\input{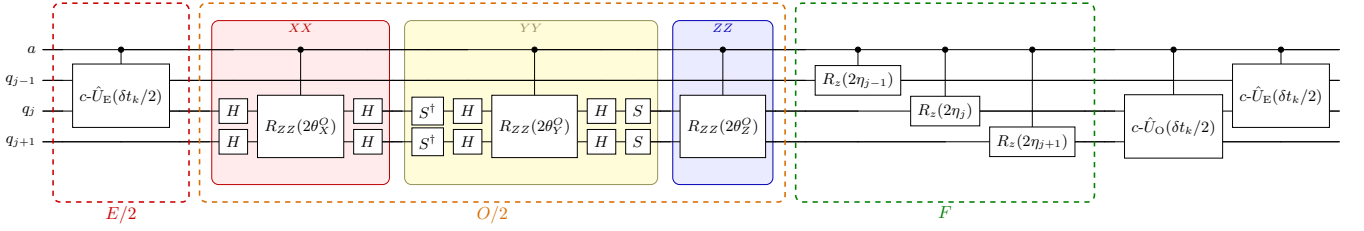}}
  \caption[RZZ-native U_RTE block]{%
  \textbf{RZZ-native implementation of the controlled real-time-evolution block used in the PITE cost evaluation.}
  The circuit is restricted to an ancilla $a$ and the three-site fragment $(q_{j-1},q_j,q_{j+1})$. One Trotter slice is the second-order Strang decomposition $c\text{-}\hat U_{\rm E}(\delta t_k/2)\,c\text{-}\hat U_{\rm O}(\delta t_k/2)\,c\text{-}\hat U_{\rm F}(\delta t_k)\,c\text{-}\hat U_{\rm O}(\delta t_k/2)\,c\text{-}\hat U_{\rm E}(\delta t_k/2)$, and $c\text{-}\hat U_{\rm RTE}(t_k)$ is obtained by $r_k$ repetitions. The first $O/2$ block is fully expanded to show how the $XX$, $YY$, and $ZZ$ interactions are realized through basis changes and ancilla-controlled $R_{ZZ}$ rotations.
  }
  \label{fig:urte}
\end{figure*}

This appendix collects the one-step PITE circuit and the RZZ-native implementation of $\hat U_{\rm RTE}$ used in the PITE resource evaluation. The $W$ gate on the ancilla line of the PITE circuit in Fig.~\ref{fig:pite_circuit} is
\begin{equation}
  W=\frac{1}{\sqrt{2}}
  \begin{pmatrix}
    1 & -i\\
    1 & i
  \end{pmatrix} .
  \label{eq:app_W_gate}
\end{equation}
The effective angle in $R_z(-2\theta_{\rm eff})$ is
\begin{equation}
  \theta_{\rm eff}
  =
  \kappa\Theta
  +s\Delta\tau_k E_{\rm shift}
  +\frac{\pi}{2}-\arctan s ,
  \label{eq:app_theta_eff}
\end{equation}
with
\begin{equation}
  s=\frac{m_0}{\sqrt{1-m_0^2}} ,
  \qquad
  \Theta=\arccos\!\left(\frac{m_0+\sqrt{1-m_0^2}}{\sqrt{2}}\right) ,
  \label{eq:app_s_theta}
\end{equation}
and
\begin{equation}
  \kappa=
  \begin{cases}
  +1, & m_0\ge 1/\sqrt{2} ,\\
  -1, & m_0< 1/\sqrt{2} .
  \end{cases}
  \label{eq:app_kappa}
\end{equation}
$E_{\rm shift}$ is the DMRG-informed energy shift used in the main text, and $\alpha=s\Delta\tau_k$. Figure~\ref{fig:urte} gives the RZZ-native implementation of the controlled real-time-evolution block used in the PITE cost evaluation.

\section{\label{app:fit_windows}Tail-fit window diagnostics}

In Fig.~\ref{fig:encoding_barrier}, we fit the post-$\Lstar$ tail of the per-site infidelity as
\begin{equation}
  \frac{1-F(l)}{N}=C(N)+A(N)e^{-k(N)l} .
  \label{eq:app_tail_fit}
\end{equation}
The fitting window is selected manually for each $N$. The selection criteria are: the window must lie beyond $\Lstar$, exhibit approximately exponential decay on a semi-log plot, end before the regime where the numerical or saturation floor begins to dominate, and contain enough points for a stable three-parameter fit. The chosen windows are listed in Table~\ref{tab:tail_fit_windows}. Within each window we perform constrained nonlinear least-squares fits subject to $C,A,k\ge0$.

\begin{table}[h]
  \centering
  \caption[Tail-fit windows]{Layer windows used for the tail fits in Eq.~\eqref{eq:app_tail_fit}.}
  \label{tab:tail_fit_windows}
  \begin{ruledtabular}
  \begin{tabular}{cc}
  $N$ & fitting window in layer $l$ \\
  \hline
   8 & $20$--$100$ \\
  10 & $50$--$250$ \\
  12 & $200$--$900$ \\
  14 & $100$--$400$ \\
  16 & $1100$--$1800$ \\
  18 & $200$--$400$ \\
  20 & $1100$--$1800$ \\
  \end{tabular}
  \end{ruledtabular}
\end{table}

\begin{figure*}[ht]
  \centering
  \includegraphics[width=\textwidth]{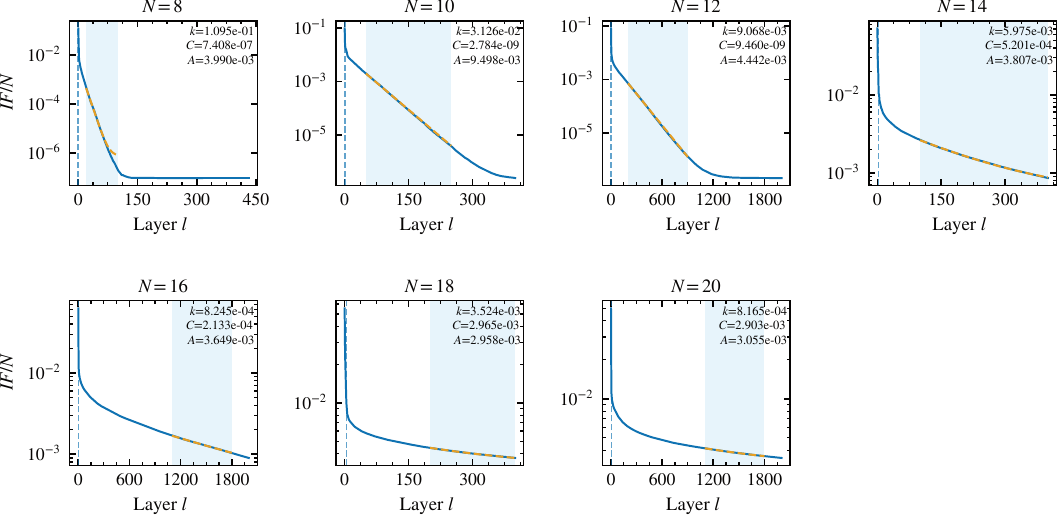}
  \caption[Tail-fit window diagnostics for hz=0]{%
  \textbf{Tail fit for $h_z=0$.}
  Shaded regions: the fitting windows of Table~\ref{tab:tail_fit_windows}. They are used to extract $C(N)$, $A(N)$, and $k(N)$ in Eq.~\eqref{eq:app_tail_fit}.
  }
  \label{fig:app_tail_fit_windows_hz0}
\end{figure*}

\begin{figure*}[ht]
  \centering
  \includegraphics[width=\textwidth]{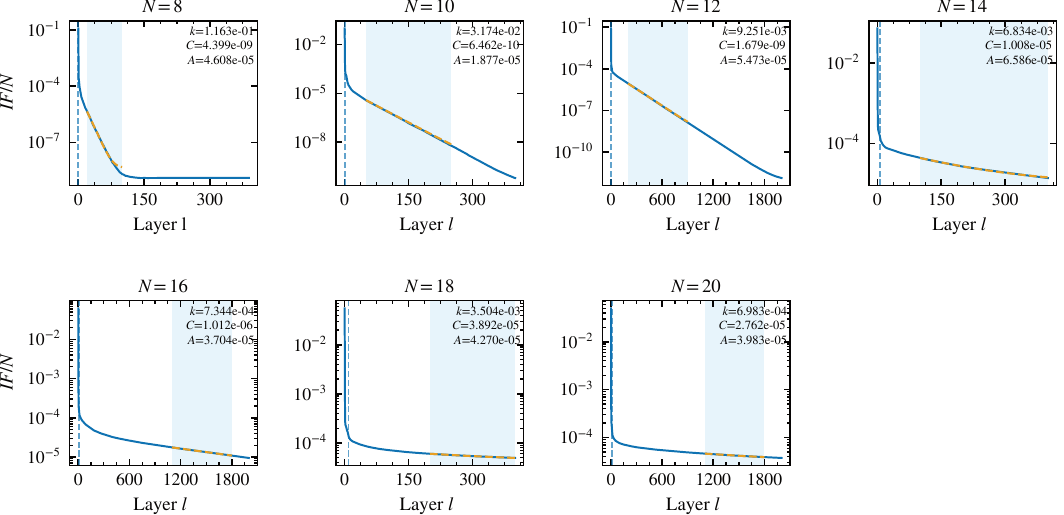}
  \caption[Tail-fit window diagnostics for hz=0.5]{%
  \textbf{Tail fit for $h_z=0.5$.}
  Shaded regions: the fitting windows of Table~\ref{tab:tail_fit_windows}. They are used to extract $C(N)$, $A(N)$, and $k(N)$ in Eq.~\eqref{eq:app_tail_fit}.
  }
  \label{fig:app_tail_fit_windows_hz0p5}
\end{figure*}

Figures~\ref{fig:app_tail_fit_windows_hz0} and \ref{fig:app_tail_fit_windows_hz0p5} show the adopted windows together with the fitted data points and curves.

\section{\label{app:tail_fit_params}Tail-fit parameters}

\begin{figure*}[ht]
  \centering
  \includegraphics[width=0.96\textwidth]{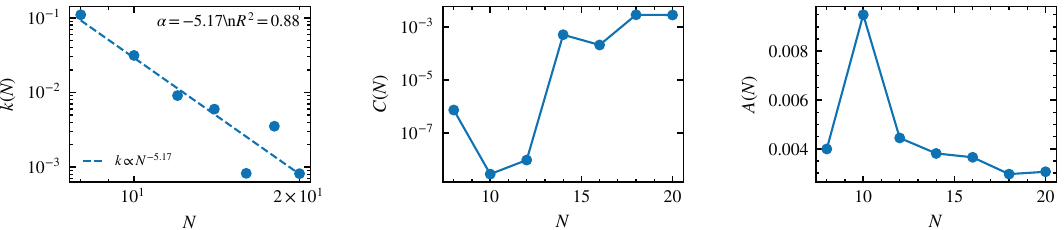}
  \caption[Tail-fit parameters for hz=0]{%
  \textbf{Tail-fit parameters for $h_z=0$.}
  Panels show $A(N)$, $k(N)$, and $C(N)$ of Eq.~\eqref{eq:app_tail_fit}.
  }
  \label{fig:app_tail_fit_params_hz0}
\end{figure*}

\begin{figure*}[ht]
  \centering
  \includegraphics[width=0.96\textwidth]{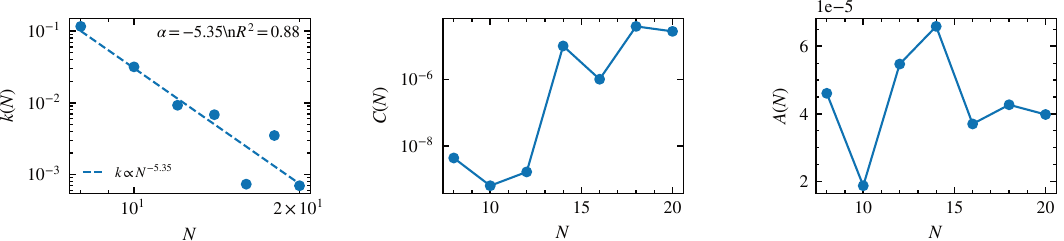}
  \caption[Tail-fit parameters for hz=0.5]{%
  \textbf{Tail-fit parameters for $h_z=0.5$.}
  Panels show $A(N)$, $k(N)$, and $C(N)$ of Eq.~\eqref{eq:app_tail_fit}.
  }
  \label{fig:app_tail_fit_params_hz0p5}
\end{figure*}

Figures~\ref{fig:app_tail_fit_params_hz0} and \ref{fig:app_tail_fit_params_hz0p5} show the system-size dependence of $A(N)$, $k(N)$, and $C(N)$ extracted from Eq.~\eqref{eq:app_tail_fit}. The constant $C(N)$ is the $\eps/N$ floor that remains at large layer counts in the encoding-only tail model. The empirical scaling $k(N)\propto N^{-5}$ quoted in the main text is the finite-size trend obtained from the windows of Table~\ref{tab:tail_fit_windows}.

\clearpage

% ============================================================

\bibliographystyle{apsrev4-2}
\bibliography{reference}

\end{document}